\documentclass[]{JFM-FLM_Au}

\usepackage{tikz}
\usepackage{bbold}

\lefttitle{X. Di and G. Benham}
\righttitle{Journal of Fluid Mechanics}

\title{Darcy plume oscillations at low injection velocities}

\author{Xiaoning Di \aff{1}   \and Graham P. Benham\aff{2} }  

\affiliation{\aff{1}School of Naval Architecture and Ocean Engineering, Dalian University of Technology, Dalian 116024, PR China
\aff{2}Department of Mechanical Engineering, University College London, London, WC1E 6BT, UK}

\corresau{Xiaoning Di, xiaoning.di@dlut.edu.cn; Graham P. Benham, g.benham@ucl.ac.uk}

\begin{document}
\maketitle

\begin{abstract}
We numerically investigate miscible buoyancy-driven Darcy plumes at low injection velocities. Below a critical injection velocity, the steady plume becomes unstable and develops periodic oscillations. Approaching the critical velocity, the oscillation amplitude vanishes while the frequency remains finite at the critical point, consistent with a supercritical Hopf bifurcation, whereas at lower injection velocities increasingly complex dynamics emerge. Unsteadiness also strongly enhances lateral mixing, increasing the mixing-layer thickness by a factor of approximately 2–3 compared with a steady plume. These findings reveal a previously unexplored dynamical regime of Darcy plumes with important consequences for subsurface transport and leakage.
\end{abstract}



\section{Introduction}

Buoyancy-driven flow in porous media arises in many geophysical and environmental applications, including geological carbon sequestration, contaminant transport in aquifers, saltwater intrusion, and geothermal heating systems \citep{Huppert2014,Woods2015,hinton2018buoyancy,benham2021two,di2025low,mondal2019modelling, bearbook}. When a lighter fluid is injected vertically into a porous medium saturated with a denser ambient fluid, the density difference generates buoyancy forces that drive the upward migration of the injected fluid, forming a buoyancy-driven Darcy plume \citep{lyu2016experimental,hewitt2020buoyancy,benham2023near, bonazzi2023influence, panda2025experimental}. Meanwhile, during transport downstream, lateral diffusion and mixing occur, causing the plume to widen and dilute \citep{sahu2015filling,gilmore2022leakage}. 

In miscible systems, local changes in concentration lead to changes in density, which in turn alter the local buoyancy and the Darcy velocity. In this coupled flow–concentration system, the natural buoyancy velocity scale, $w_b$, can be derived from Darcy’s law, namely
\begin{equation}
    w_b
    =
    \frac{K\Delta\rho g}{\mu},\label{buoyw}
\end{equation}
where $K$ is the permeability of the porous medium, $\Delta\rho>0$ is the density difference between the fluids, $\mu$ is the dynamic viscosity, and ${g}$ is gravitational acceleration. 

Motivated by understanding leakage dynamics near geological faults during CO$_2$ sequestration, as well as other environmental leakage flows, such as contaminant release in aquifers \citep{bearbook,Huppert2014,gilmore2022leakage,benham2023prsa,benham2023near,di2026propagation}, we focus on fluid injection from a constant flux point source. Within this context, a key parameter governing the problem is the ratio $W$ between the injection velocity at the source $w_0$ and the buoyancy velocity Eq. ~\eqref{buoyw} \citep{benham2023near}, such that
\begin{equation}
    W
    =
    \frac{w_0}{w_b}.
\end{equation}
This quantifies the relative strength of the imposed source compared to the buoyancy. As described by \citet{benham2023near}, the near-field shape of the plume (in the case of a macroscopic sharp interface) is fully determined by $W$, with $W<1$ leading to locally thinning plumes, and $W>1$ to locally thickening plumes, with steady plume shapes found for $W$ values in the range $0.12-1.14$. 

The size of $w_0$ and hence $W$ is determined by the properties of the inlet, such as the fault geometry and permeability \citep{gilmore2022leakage}. In the case where the driving pressure head is small compared with the pressure head across the fault, or when the fault permeability is small compared to the reservoir permeability, this leads to small values of $W\ll1$. Such cases are associated with plumes that  thin dramatically near the inlet, causing large, negative, flow-enhancing pressures locally that play a significant role on leakage dynamics. However, it was shown by \citet{benham2023near} that below a critical value $W<W_c\approx 0.5$, steady plume shapes become linearly unstable due to the fact that the steep plume shape cannot support the unstable density difference locally (essentially a Rayleigh-Taylor instability). In addition, there is a lower critical value $W_0=0.12$, below which no steady plume shape appears to exist, raising questions about the fate of the Darcy plume for such small inlet velocities.

However, beyond predicting the instability and the disappearance of steady solutions at sufficiently low \(W\), this has never been explored further, nor is it known what occurs after the onset of the instability.  
Hence, for small $W$ values, the long-time evolution of the plume shape remains unclear, as well as the impact on concentration transport and mixing with the ambient fluid.

To address the above issues, this study uses numerical simulations to investigate the unsteady behavior of buoyancy-driven miscible Darcy plumes at low inlet velocities. The reliability of the numerical model is verified by comparing it with experimental results and theoretical solutions for steady states and then extended to simulate unsteady states. We find that the unsteady Darcy plume can enter a periodic oscillation (or limit cycle) with a characteristic frequency and an amplitude that diminishes as the critical value $W_c$ is reached, which is consistent with a supercritical Hopf bifurcation, similar to the case of a dripping faucet \citep{pinto1995hopf}.

\section{Numerical model}
We numerically investigate the buoyancy-driven Darcy plume in a saturated two-dimensional porous medium. The governing equations consist of the continuity equation, Darcy’s law, and the advection-diffusion equation, given by
\begin{gather}
\nabla\cdot\mathbf{U}=0,\\
\mathbf{U}
=
-\frac{K}{\mu}
\left(
\nabla P+\rho(c)g\mathbf{e}_z
\right),\\
\phi\frac{\partial c}{\partial t}
+
\nabla\cdot(c\mathbf U)
=
\phi D\nabla^2c.
\end{gather}
Here, $\mathbf{U}(x,z,t)$ is the Darcy velocity, $P(x,z,t)$ is the pressure, $c(x,z,t)$ is the concentration, $\phi$ is the porosity, and $D$ is the diffusion coefficient. The viscosity $\mu$ is assumed to be independent of concentration. The density $\rho(c)$ is assumed to vary linearly with concentration, written as
\begin{equation}
    \rho(c)
    =
    \rho_l
    +
    \left(\rho_h-\rho_l\right)c,
\end{equation}
where $\rho_l$ and $\rho_h$ are the constant densities of the light and dense fluids. Therefore, $c=0$ denotes the light fluid and $c=1$ the dense fluid. 
We ignore the effects of mechanical dispersion for simplicity, although these could be explored by adding an effective diffusion tensor that depends on the local flow velocity \citep{Woods2015}.

We implement the governing equations using a finite-volume formulation in the open-source computational-fluid-dynamics platform OpenFOAM\textsuperscript{\textregistered} \citep{weller1998tensorial}. 
The computational domain is discretized using approximately $5\times10^{5}$ cells, with local mesh refinement near the plume inlet and the interfacial region between two fluids. The pressure and concentration equations are solved to absolute tolerances of $10^{-8}$ and $10^{-9}$, respectively. Utilizing symmetry about the vertical axis, only half of a plume is investigated within a square computational region ($L_x\times L_z=$ 0.4 m $\times$ 0.4 m). We follow the experimental parameters of \citet{gilmore2022leakage}, in which salty water is injected into the top of a reservoir of porous beads saturated with fresh water. In those experiments, the salty water forms a dense plume that sinks downwards. Since we are interested in rising plumes (which are more relevant for most environmental applications), we reverse the fluids, such that buoyant fresh water is injected into the bottom of a reservoir filled with salty water. Under the Boussinesq approximation, the flow patterns should remain equivalent despite the vertical flipping \citep{amooie2018solutal}. As described by \citet{gilmore2022leakage}, the permeability and porosity of the porous beads were estimated as $K=1.06 \times 10^{-8}$ m$^{2}$ and $\phi=0.41$. 

\begin{figure}
	\centering
\includegraphics[width=0.88\columnwidth]{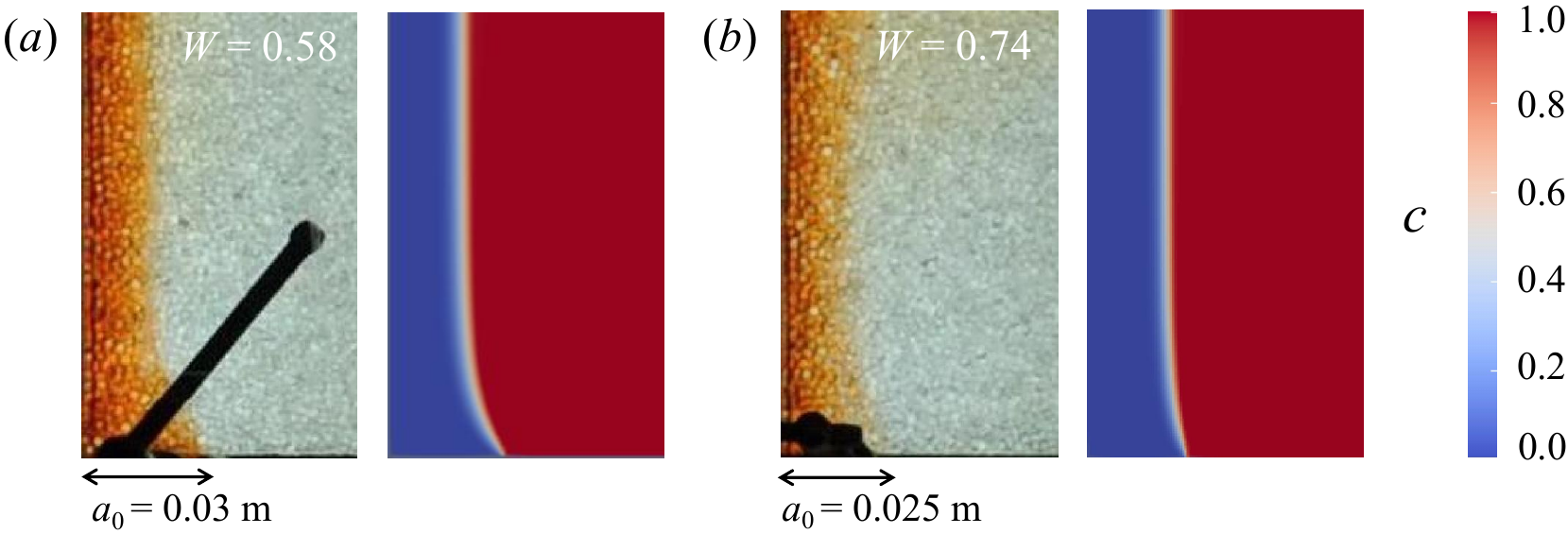}
    \caption{Validation of the numerical model for steady plume shapes. (a,b) Comparison between experimental observations of fresh and salty water in a tank of porous beads (Gilmore et al. 2022) and numerical simulations for $W=0.58$ and $W=0.74$. Experimental photos are partially obstructed by a clamp.}
	\label{experiment}
\end{figure}

The chosen idealised boundary conditions are summarised as follows. The left side $x=0$ is set as a symmetry boundary, while the top and right boundaries ($z=L_z$, $x=L_x$) are set as outlets with hydrostatic pressure. On the bottom boundary $z=0$, a velocity inlet of width $a_0=0.03$ m is located at the lower-left corner, and the remainder of the bottom boundary is impermeable. Initially, the  domain is saturated  with saline water ($\rho_h=1030$ kg/m$^{3}$; $\mu=0.001$ Pa$\cdot$s). Fresh water with lower density and the same viscosity ($\rho_l=998$ kg/m$^{3}$; $\mu=0.001$ Pa$\cdot$s) is then injected through the inlet at constant vertical velocity $w_0$. As such, the inlet buoyancy velocity (before mixing) is estimated as $w_b=3.33\times 10^{-3}$ m/s.   
We take the molecular diffusivity of salt in water as $D=1.5\times 10^{-9}$ m$^2$/s. Hence, defining the P\'{e}clet number as Pe $=w_b a_0/D$, we find Pe $\approx 6.6\times 10^4$ (although a range of Pe values are studied later in Section \ref{sec_bif}).
Based on this numerical configuration, we vary $W$ via $w_0$ to investigate the dynamics of the Darcy plume.

\section{Results}

\subsection{Comparison with experiments and theory for steady states}

\begin{figure}
	\centering
\includegraphics[width=0.88\columnwidth]{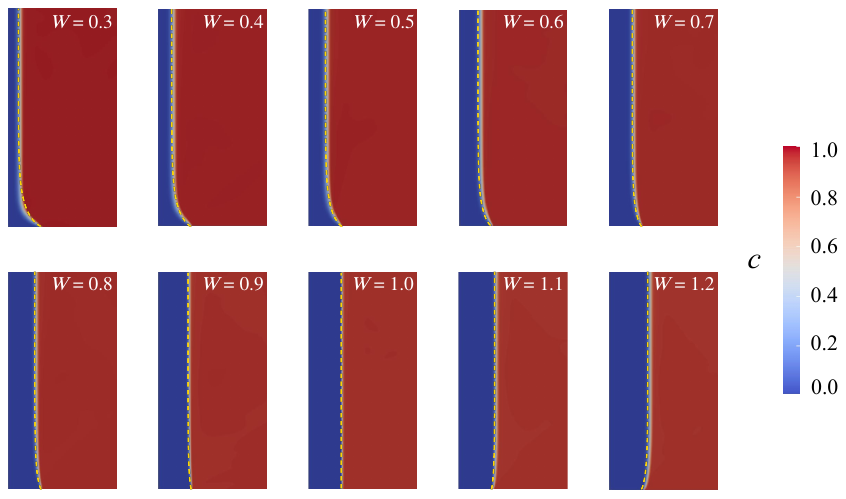}
    \caption{Comparison between theoretical solutions \eqref{eq:plume} (yellow lines) and numerical simulations for intermediate values of $W$. }
	\label{validation}
\end{figure}

We begin by testing the validity of the numerical model. We first compare the results of numerical simulations with the steady Darcy plume observed in experiments of salty and fresh water in a tank of porous beads \citep{gilmore2022leakage}, as shown in figure \ref{experiment}. As can be seen in figure \ref{experiment}(a,b), for both cases ($W=0.58$ and $W=0.74$), the numerical simulations successfully reproduce the experimental plume shape, and the near-field regions of the plumes are fundamentally consistent. We further compare the numerical simulations with the theoretical solution for the near-field interface of steady Darcy plumes in the absence of mixing \citep{benham2023near}. In the case where $W\approx 1$ it was shown that the plume half-width $a(z)$ approximately satisfies the differential equation
\begin{equation}
    a'(z) \approx -(1-W)\frac{4}{\pi}
    \tanh^{-1}\!\left[
        \exp\left(-\frac{\pi z}{2a_0}\right)
    \right],
    \label{eq:plume}
\end{equation}
where $a_0=a(0)$ is the inlet half-width. 
Equation \eqref{eq:plume} predicts locally widening, straight and narrowing plumes for \(W>1\), \(W=1\) and \(W<1\), respectively. 
As shown in figure \ref{validation}, the numerical results reproduce the theoretical near-field plume shapes fairly well, providing a basis for investigating the unsteady dynamics at lower $W$.


\subsection{Unsteady near-field dynamics}
\begin{figure}
    \centering
    \includegraphics[width=0.88\columnwidth]{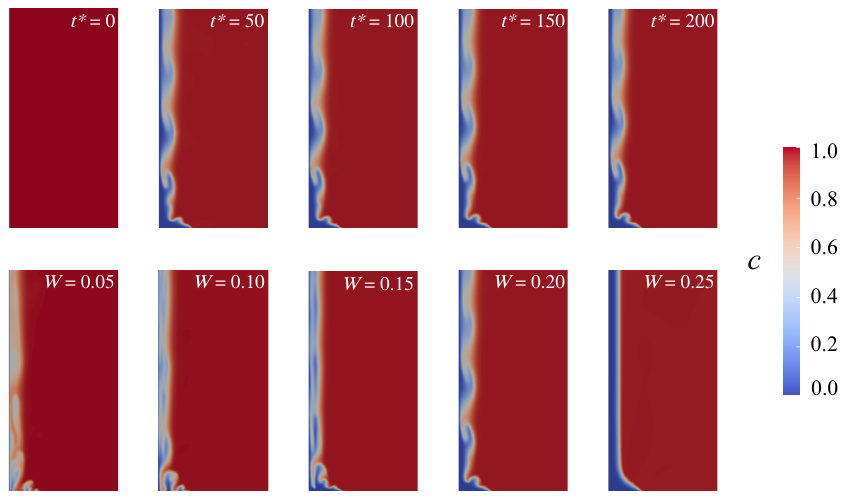}
    \caption{Numerical observations of unsteady Darcy plumes for small $W$.
    First row: evolution of Darcy plume at $W=0.2$;
    Second row: Comparison of Darcy plumes under low $W$ conditions at $t^*$ = 200.}
    \label{unsteady}
\end{figure}

\begin{figure}
    \centering
    \begin{tikzpicture}
        \node at (0,0) {\includegraphics[width=0.88\columnwidth]{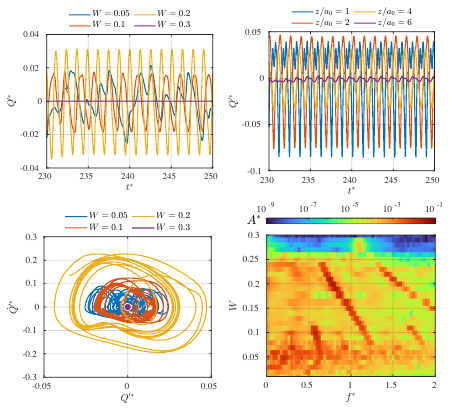}};
        \node at (-5.6,4) {(a)};
        \node at (0,4) {(b)};
        \node at (-5.6,-1) {(c)};
        \node at (0,-1) {(d)};
    \end{tikzpicture}
\caption{
Unsteady plume dynamics in the near field.
(a) Time series of the dimensionless flux fluctuation $Q^{\prime *}$ at $z/a_0=4$ for representative values of $W$.
(b) Time series of $Q^{\prime *}$ at different downstream locations ($z/a_0=1,2,4,$ and $6$) for $W=0.20$.
(c) Phase trajectories in the $(Q^{\prime *},\dot{Q}^{\prime *})$ plane at $z/a_0=4$.
(d) Distribution of the dimensionless FFT amplitude $A^*$ as a function of dimensionless frequency $f^*$ and $W$ at $z/a_0=4$.
}
    \label{dynamic}
\end{figure}

Now, we reduce $W$ to investigate the near-field dynamics as buoyancy becomes increasingly important relative to the imposed source injection. To quantify the plume dynamics, we define the plume flux across a horizontal
section as
\begin{equation}
    Q(z,t)
    =
    \int_{0}^{L_x}
    [1-c(x,z,t)]\,w(x,z,t)\,\mathrm{d}x.
    \label{eq:plume_flux}
\end{equation}
The corresponding flux fluctuation is
\begin{equation}
    Q'(z,t)
    =
    Q(z,t)-\overline{Q(z,t)},
    \label{eq:Qprime}
\end{equation}
where the overbar denotes a temporal average over the developed state. 
Furthermore, we introduce the following dimensionless quantities
\begin{equation}
    Q^{\prime *}=\frac{Q^\prime}{a_0 w_b},
    \quad
    t^*=\frac{t w_b}{a_0},
    \quad
    \dot{Q}^{\prime *}=\frac{\dot{Q}^\prime}{w_b^2},
    \quad
    f^*=\frac{f a_0}{w_b},
\end{equation}
where $f$ is the frequency of oscillation, and dots denote time derivatives.

Figure \ref{unsteady} shows the transition from strongly unsteady plumes at low \(W\) to steady plumes above a critical value \(W_c\approx0.25\). At low \(W\), near-source fluctuations strongly deform the plume and enhance mixing, whereas above \(W_c\) a stable plume forms with mixing confined to a thin interfacial layer.


The critical value \(W_c\approx0.25\) is smaller than the sharp-interface prediction \(W_c\approx0.5\) of \citet{benham2023near}, suggesting that diffusion may have a stabilising effect on the strongly thinned near-source plume (in fact, later in Section \ref{sec_bif} we show how $W_c$ varies with Pe). The predicted disappearance of unstable steady states below \(W_0\approx0.12\) cannot be directly assessed here, since such states are not realised dynamically.


Figure \ref{dynamic}a shows a transition from steady behaviour at \(W=0.30\) to coherent oscillations at \(W=0.20\) and \(0.10\), and increasingly irregular oscillations at \(W=0.05\).
Figure \ref{dynamic}b shows that the oscillations have approximately the same period throughout the plume, while their amplitude is largest near the inlet and decays downstream, demonstrating spatially coherent oscillatory behaviour.

The phase portraits in figure \ref{dynamic}c further demonstrate the transition from a steady state to closed periodic trajectories, which become increasingly irregular at low \(W\). Correspondingly, the dimensionless spectral amplitude $A^*$ in figure \ref{dynamic}d (obtained via Fast Fourier Transform) shows a distinct dominant frequency $f^*$ for \(0.06\lesssim W<W_c\), together with higher harmonics. For \(W<0.06\), the spectrum broadens, indicating increasingly complex dynamics, whereas for \(W>W_c\) the coherent spectral features disappear.

The periodic behaviour can be interpreted as a repeated cycle of plume thinning and Rayleigh–Taylor instability near the source. At low injection velocities, buoyancy strongly narrows the plume, producing a locally unstable density configuration that deforms the plume and releases a buoyant parcel of fluid upwards. This deformation temporarily relaxes the unstable configuration, before continued injection re-establishes the narrow plume and the cycle repeats.

\subsection{Unsteady near-field lateral mixing}

Next, we examine how the unsteady Darcy plume affects the near-field mixing. For a steady plume, the concentration distribution on the plume interface can be approximated by a hyperbolic tangent function
\begin{equation}
    c(x,z)
    =
    \frac{1}{2}
    \left[
        1+
        \tanh\left(
            \frac{x-x_c(z)}{\delta(z)}
        \right)
    \right],
    \label{eq:tanh_profile}
\end{equation}
where $x_c(z)$ denotes the centre of the concentration transition and $\delta(z)$ is its characteristic mixing-layer thickness. The parameters $x_c$ and $\delta$ are obtained by fitting Eq.~\eqref{eq:tanh_profile} to the numerical concentration profile using a least-squares method at a given $z$ value.

\begin{figure}
	\centering
    \begin{tikzpicture}
        \node at (0,0) {\includegraphics[width=0.82\columnwidth]{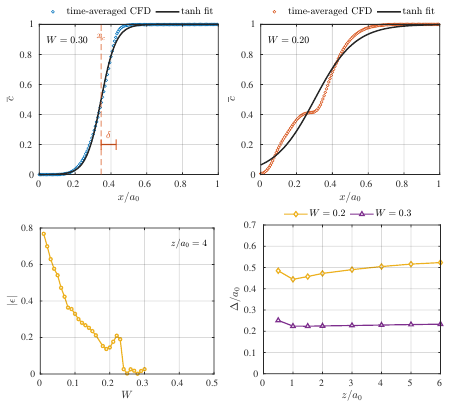}};
        \node at (-5.4,4) {(a)};
        \node at (0.1,4) {(b)};
        \node at (-5.4,-1) {(c)};
        \node at (0.1,-1) {(d)};
            \end{tikzpicture}
\caption{
Unsteady near-field mixing.
(a) Concentration profile for the steady plume at $W=0.3$ and $z/a_0=4$, together with the hyperbolic-tangent fit. The fitted centre $x_c$ and characteristic thickness $\delta$ are indicated.
(b) Concentration profile and hyperbolic-tangent fit for unsteady plume at $W=0.2$.
(c) Magnitude of the parameter $|\epsilon|$ \eqref{eq:epsilon} as a function of $W$ at $z/a_0=4$, quantifying the departure of the time-averaged concentration profile from the hyperbolic-tangent representation.
(d) Downstream evolution of the alternative mixing-layer thickness $\Delta/a_0$ for unsteady ($W=0.2$) and steady ($W=0.3$) plumes.
}
\label{mixing}
\end{figure}

Figure ~\ref{mixing}a shows the time-averaged concentration profile of a steady plume under the conditions $W=0.30$ and $z/a_0=4$. The numerical profile is well described by Eq.~\eqref{eq:tanh_profile}, as shown by the fit. By contrast, in the unsteady case shown in figure ~\ref{mixing}b, where $W=0.20$, the concentration distribution exhibits a significant non-monotonic deformation, which cannot be described by \eqref{eq:tanh_profile}. 

To quantify this deviation from the hyperbolic tangent distribution, we define
\begin{equation}
    \epsilon(z)
    =
    \frac{1}{2\delta(z)}
    \int_0^{L_x}4\bar{c}(x,z)(1-\bar{c}(x,z))\,\mathrm{d}x
    -1,
    \label{eq:epsilon}
\end{equation}
where $\bar{c}$ denotes the time-averaged concentration. For the hyperbolic tangent distribution, $\epsilon=0$; whereas an increase in $|\epsilon|$ indicates a larger deviation from the prescribed distribution. Figure ~\ref{mixing}c shows that when $W$ is larger than the critical value $W_c$, $|\epsilon|$ remains small; however, as $W$ decreases below $W_c$, leading to unsteady states, $|\epsilon|$ increases markedly. Note that to calculate $\epsilon$, $\delta$ is obtained from the best-fit profile even when the concentration distribution deviates from \eqref{eq:tanh_profile}. 

We define an alternative mixing-layer thickness $\Delta(z)$ as the difference between the $5\%$ and $95\%$ time-averaged concentration values, such that
\begin{equation}
    \Delta(z)
    =
    x(\bar{c}=0.95)-x(\bar{c}=0.05).
\end{equation}
Figure ~\ref{mixing}d compares the downstream evolution of $\Delta(z)$ for $W=0.20$ and $0.30$. For the steady plume at $W=0.30$, $\Delta/a_0$
remains at a relatively small value and is almost independent of the downstream distance. In contrast, the unsteady plume at $W=0.20$ exhibits a significantly wider mixing layer (by a factor of approximately  2-3), which gradually widens downstream. 
Thus, the unsteady plume exhibits significantly enhanced lateral mixing, with interface deformation promoting mixing by molecular diffusion.

\subsection{Bifurcation behaviour and parameter dependence}

\label{sec_bif}

\begin{figure}
	\centering
	    \begin{tikzpicture}
	\node at (0.3,0) {\includegraphics[width=0.85\columnwidth]{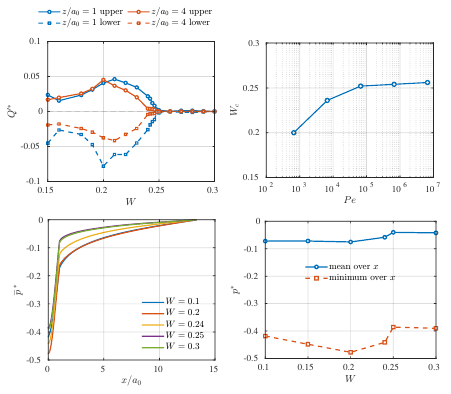}};
	\node at (-5.05,4) {(a)};
	\node at (0.6,4) {(b)};
	\node at (-5.05,-0.5) {(c)};
	\node at (0.6,-0.5) {(d)};
	\end{tikzpicture}
\caption{
 Bifurcation behaviour and near-field pressure. (a) Upper and lower envelopes of the dimensionless flux fluctuation $Q'^*$ as functions of $W$, evaluated at $z/a_0=1,4$. (b) Critical injection velocity ratio $W_c$ as a function of the Péclet number Pe at $z/a_0=4$. (c) Dimensionless pressure $p^*$ at $z=0$ for different values of $W$. (d) Mean and minimum dimensionless pressure as functions of $W$.
}
\label{Fig:6}
\end{figure}

The above results indicate that near-field Darcy plumes maintain oscillatory behaviour for a range of $W$ values. We now identify the bifurcation behaviour between unsteady and steady states near the critical value $W\approx W_c$. 
Figure \ref{Fig:6}a shows the oscillation envelopes of \(Q'^{*}\) at several downstream locations. As \(W\to W_c^-\), the oscillation amplitude vanishes at \(W_c\approx0.25\), while the dominant frequency remains finite (\(f^*=0.65\)–\(0.68\)), which is consistent with a supercritical Hopf bifurcation.
This behaviour is reminiscent of the dripping-faucet problem, which similarly exhibits a Hopf bifurcation, where a steady stream switches to periodic dripping and eventually chaos as the flow rate is reduced \citep{pinto1995hopf,coullet2005hydrodynamical,subramani2006simplicity}.


Figure \ref{Fig:6}b  shows the influence of molecular diffusion on the instability via the P\'{e}clet number. As Pe increases, $W_c$ increases towards a diffusion-independent limit, while at sufficiently small Pe the oscillations disappear altogether. Thus, diffusion stabilises the plume and partly accounts for the difference between the present critical value and the sharp-interface prediction \citep{benham2023near}.

Finally, figures \ref{Fig:6}c,d show the dimensionless pressure near the inlet ($p^*=p/\Delta\rho g a_0$ at $z=0$) associated with the strongly thinning near-field plume. The pressure decreases sharply near the inlet, producing a pronounced local pressure minimum for small $W$. This negative pressure is a consequence of the buoyancy-driven acceleration of the thinning plume and has important implications for leakage flows, as discussed below.

\section{Conclusions}

This study has revealed a previously unexplored unsteady regime of miscible buoyancy-driven Darcy plumes at low injection velocities. Below a critical value $W_c$, the steady plume gives way to periodic oscillations associated with repeated near-source thinning and Rayleigh–Taylor instability. The oscillation amplitude vanishes while the dominant frequency remains finite as $W\rightarrow W_c^{-}$, consistent with a supercritical Hopf bifurcation, while increasingly complex dynamics emerge at still lower $W$. The critical value depends on molecular diffusion: $W_c$ increases with Pe before approaching an approximately constant value in the weak-diffusion limit.

These unsteady dynamics have important consequences for subsurface transport. Plume oscillations strongly deform the concentration field and enhance lateral mixing, producing a mixing layer approximately 2–3 times thicker than that of a steady plume. In addition, the strongly thinning near-source plume generates a pronounced negative pressure near the inlet. For leakage through a permeable fault or pathway whose flux is coupled to the local pressure difference, this negative pressure would act to enhance the driving pressure and hence the leakage rate \citep{gilmore2022leakage,benham2023near}. Thus, the onset of plume instability can affect not only the subsequent mixing and transport of leaked fluid, but also the leakage rate itself.

The present model neglects mechanical dispersion in order to isolate the effects of molecular diffusion. Future work should examine velocity-dependent dispersion and three-dimensional geometries, both of which may modify the critical conditions and nonlinear plume dynamics. Extensions to immiscible flows would additionally introduce capillary pressure and relative permeability, which may modify the oscillatory instability and plume breakup, with consequences for the migration and leakage of subsurface CO$_2$ during geological carbon sequestration \citep{Huppert2014}.
\\
\\
\textbf{Funding.} X.D. acknowledges support from National Natural Science Foundation of China (Grant No. 52501324) and the Fundamental Research Funds for the Central Universities.
\\
\\
\textbf{Declaration of interests.} The authors report no conflict of interest.
\\
\\
\textbf{Author ORCIDs.}\\
Xiaoning Di         https://orcid.org/0009-0005-6433-6149\\
Graham P. Benham    https://orcid.org/0000-0003-1664-6976
\\
\\
\textbf{Use of AI tools.}\\ ChatGPT (GPT-5.6 Sol, OpenAI) was used during 2026 for copy-editing and language refinement of the manuscript. No scientific results, analyses or figures were generated or modified using AI tools. The authors have reviewed and verified the manuscript and take full responsibility for its content.
 
\bibliographystyle{jfm}
\bibliography{jfm}


\end{document}